\documentclass[12pt]{article}          
\usepackage{graphicx,epsf,subfigure,pstricks,pst-node,psfrag,amsthm,amssymb,amsmath,url,inputenc}
\begin{document}
\title{\huge \textbf{Incremental Cost-Effectiveness Statistical Inference: Calculations and Communications.}}
\author{Robert L. Obenchain \\ Risk Benefit Statistics, Clayton, CA 94517}
\date{December 2023}
\maketitle

\begin{abstract}
{\noindent We illustrate use of \textit{nonparametric statistical methods} to compare alternative treatments for
a particular disease or condition on both their relative effectiveness and their relative cost. These
\textit{Incremental Cost Effectiveness} (ICE) methods are based upon \textit{Bootstrapping} (i.e. Sampling With-Replacement)
using observational or clinical-trial data on individual patients. We first show how a reasonable numerical value for
$\lambda$, the \textit{Shadow Price of Health}, can be chosen using functions within the \textit{ICEinfer} R-package when
effectiveness is not measured in ``QALY''s. We also argue that simple histograms are ideal for communicating key ICE findings to
regulators, while our more detailed graphics may well be more informative and compelling for some health-care stakeholders.}
\end{abstract}

\section{Introduction}

We outline and update statistical methodology needed by Health-Outcomes researchers to make powerful and robust
inferences using either clinical trial or real-world (observational) data to make realistic Head-to-Head comparisons
of cost and effectiveness differences between a pair of alternative treatments for a given disease or health-care
condition. We also review and illustrate use of Incremental Cost-Effectiveness ``ICE Preference Maps'' that can be
distinctly non-linear to incorporate realistic consumer preferences from empirical health care research. 

\subsection{The Cost-Effectiveness Plane}

As proposed in Black (1990), the \textit{Cost-Effectiveness} plane displays measures of treatment ``effectiveness''
along its horizontal $x-$axis and measures of treatment ``cost'' along its vertical $y-$axis. This display is ``Incremental''
in the sense that \textit{differences in treatment outcomes} : [``New'' treatment minus ``Standard'' treatment]
are plotted. In other words, the ``ICE Origin'' at $(0,0)$ represents the relative position of the current Standard treatment.
The New treatment is then relatively \textit{More Effective} at points with $x-$values that are strictly positive and
\textit{Less Costly} at points with $y-$values that are strictly negative; see Figure $1$ on page $3$.

\begin{figure}
\center{\includegraphics[width=7in]{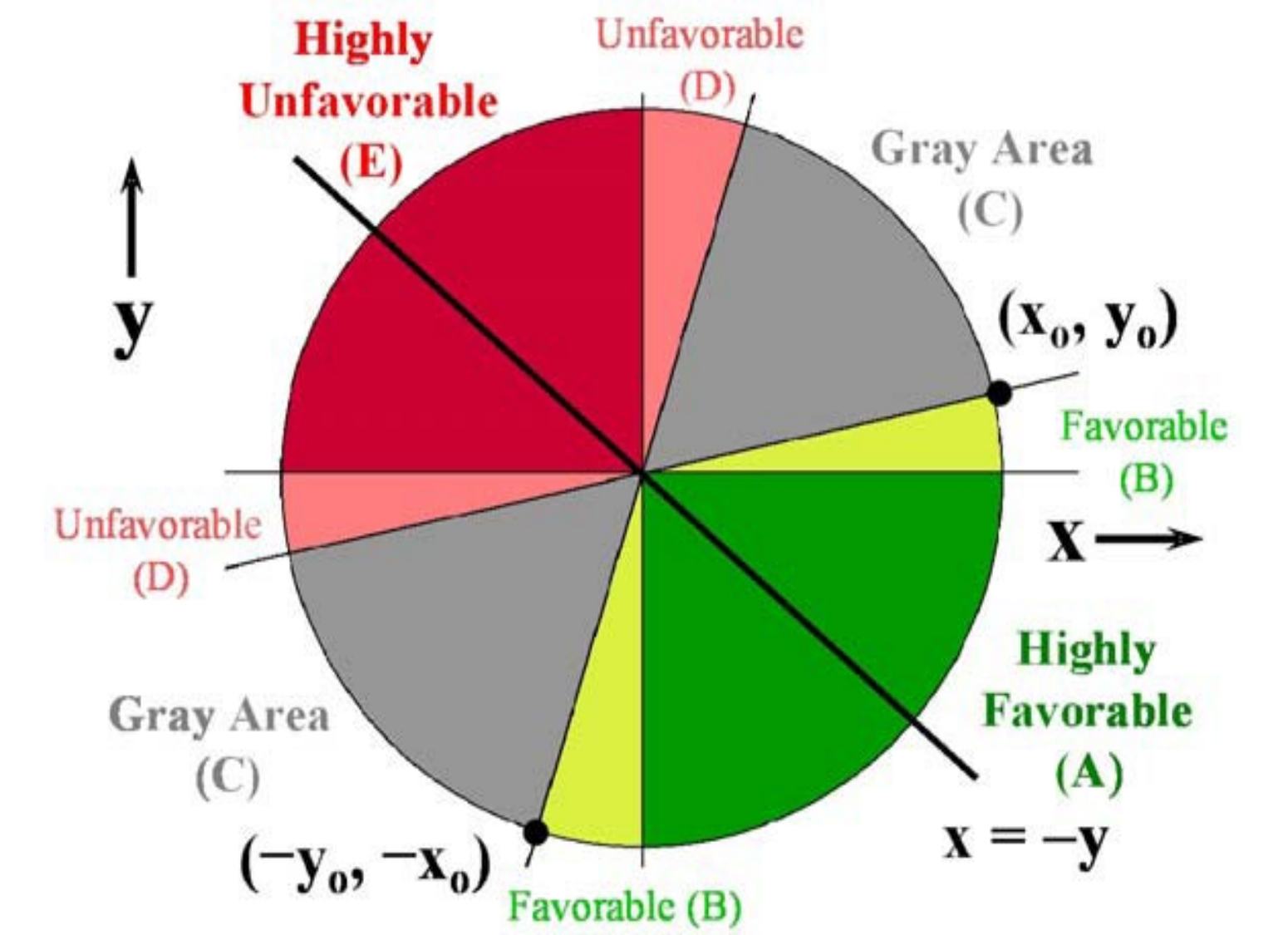}} 
\caption{\label{fig:PIE} This graphic from Obenchain (2008) illustrates some important characteristics of coherent
ICE preferences on the (Euclidean) Cost-Effectiveness plane of Black (1990). In particular, we note that preferences
are \textbf{symmetric} relative to the \textit{upper-left to lower-right diagonal line}, $x = -y$. This symmetry was
first depicted graphically in Laupacis et al.(1992) and is dictated by the Fourth Axiom of Table $1$.}
\end{figure}

Four key axiomatic properties of ICE preference maps are listed below in Table $1$. When first examining this table, it may
be helpful to note that the linear ``Net Benefit'' preference map, $NB(x,y) = x - y$ of Stinnett and Mullahy (1998)
satisfies these axioms. Subsections $\S 3.1$ through $\S 3.5$ of Obenchain (2008) provide highly detailed motivations for all
four of these axioms.

\begin{table}[ht]
\begin{tabular}{@{}cc@{}}
\multicolumn{2}{c} {\textbf{TABLE 1 -- Four Axioms of Coherent ICE Preferences}} \\
  & \\
Indifference and & $P(x,y)=0$ when $x=y$, \\
Direction of Preference & $P(x,y)>0$ when $x>y$, and $P(x,y)<0$ when $x<y$. \\
  & \\
Monotonicity & $P(x,y) \ge P(x_o,y_o)$ for all $x \ge x_o$ and $y \le y_o$. \\
  & \\
Re-labeling & $P(x,y) = -P(-x,-y)$. \\
  & \\
Symmetry and Anti-symmetry & $P(x,y) = P(-y,-x) = -P(y,x)$. \\    
\end{tabular}
\end{table}
\vspace{0.5cm}

\subsection{ICE Statistical Inference: Basics}

In Incremental Cost-Effectiveness inferences, Costs are always expressed in monetary units while Effectiveness
is measured in units considered fully appropriate for the condition or disease being treated. The \textit{ICEscale()} function
within the \textit{ICEinfer R-package}, Obenchain (2007-2020), facilitates choice of a numerical value for $\lambda$ by 
calculating a purely ``statistical'' value. This value is the ratio of the Standard Deviations of the observed
Cost and Effectiveness Differences. In the numerical example introduced below in $\S1.3$, this ratio is $4.556/0.21309 = 21.381$.

The $\lambda-$values that I personally recommend for general use are \textit{Integer Powers of 10}. In these cases, only the
location of the decimal point on axis \textit{tick-mark} labels change due to choice of $\lambda$. Thus the shadow price actually
used in our numerical example will be simply $10$ ...the \textit{Integer Power of 10} closest to $21.381$.

When cost-measurements are re-expressed using a different currency and/or alternative effe-measures are reported, the
\textit{ICEscale()} function typically proposes using a different ``Integer Power of 10'' shadow price. Plots generated by functions
within the \textit{ICEinfer R-}package may also adjust the aspect ratio of a plot to make it appear more nearly ``Square''.

Some regulatory authorities and ``Health Outcomes'' researchers prefer expressing treatment Effectiveness in Quality
Adjusted Life Years (QALYs) preserved; see: Weinstein, Torrance and McGuire (2009) or Neumann, Cohen and Weinstein (2014).
Unfortunately, re-expressing effectiveness estimates in QALYs can be both difficult and \textbf{highly subjective} ...especially when the condition being treated is \textbf{not life-threatening}.

\subsection{ICE Preferences}

The ICE Preference Maps (functions) introduced in Obenchain (2008) are of the general form: 

\begin{equation}
P(x, y) \propto (x^2 + y^2)^{(\beta - \gamma)/2}[x-y]^{\gamma}, \label{eqn: Prefr}
\end{equation}

\noindent where $\propto$ means ``is proportional to'', $\beta$ and $\gamma$ are strictly positive ``power'' parameters, and
the special notation $[z]^c$ denotes a ``signed-power.'' Specifically, $[z]^c$ denotes the product of sign(z), which is +1,
0 or -1, times the absolute value of z raised to the power c. Special care is taken in expressing equation (1) because
non-integer powers of negative real numbers are generally imaginary. ICE preferences need to be expressed as real values even
though they may provide only ordinal measures of preference strength.

It is straight-forward to verify that all ICE maps of form (1) satisfy axioms 1, 3 and 4 of Table 1. The \textit{linear preference map},
$NB(x,y) = x - y$ of Stinnett and Mullahy (1998) is the special case where $\beta = \gamma = 1$.

To satisfy axiom 2, the following restriction on the ratio of the $\gamma$ and $\beta$ \textit{power parameters} is required:
\begin{equation}
0.1715729... = 1/\Omega \le {\gamma}/\beta \le \Omega = (1+\sqrt{2})^2 = 5.828427... \label{eqn: ICEomega}
\end{equation}

For a given numerical value of $\lambda$, the difference in \textit{Effectiveness} between treatments, $\Delta E$, can
be re-expressed in cost units via multiplication: $\lambda \times \Delta E$. Similarly, a difference in Cost between treatments,
$\Delta C$, would be converted into Effectiveness units by division: $\Delta C/\lambda$.

\subsection{Returns-to-Scale}

Suppose now that the observed treatment differences in cost, $y$, and effectiveness, $x$, are somehow both multiplied by a strictly positive
and finite real valued factor, $f$. In other words, the observed effectiveness difference of $x$ becomes $f*x$, while the observed cost
difference of $y$ becomes $f*y$. The resulting new value of preference in Equation $1$ is then:

\begin{equation}
P(f*x, f*y) \propto f^{(\beta - \gamma)+ \gamma}*(x^2 + y^2)^{(\beta - \gamma)/2}*[x-y]^{\gamma} \propto f^\beta * P(x,y). \label{eqn: RtS}
\end{equation}

\noindent In other words, for every map in our 2-parameter family, returns-to-scale depend solely upon the $\beta$ power parameter associated
with only the ICE radius factor. Specifically, \textit{returns-to-scale} will be:

\textbf{Decreasing} when $0 < \beta < 1$;

\textbf{Constant} [linear] when $\beta = 1$; or

\textbf{Increasing} when $\beta > 1$ and finite.

\subsection{A Numerical Example using simulated Effectiveness and Cost Data}

Table [$2$] provides Summary Statistics for the patient-level data that we will analyze to illustrate basic concepts.

\begin{table}[ht]
\begin{tabular}{@{}ccccccc@{}}
\multicolumn{7}{c} {\textbf{TABLE 2 -- Summary Statistics for the Effe-Cost Example}} \\
Variable & Min. & 1st.Qt & Median & Mean & 3rd.Qt & Max. \\  
  & & & & & & \\
99 Std. Treatment Patients & & & & & & \\ 
effe & 0.076 & 2.436 & 3.615 & 3.653 & 4.624 & 8.373 \\ 
cost & 0.288 & 53.857 & 78.073 & 76.497 & 98.436 & 172.0 \\
  & & & & & & \\
101 New Treatment Patients & & & & & & \\
effe & 0.816 & 3.210 & 3.890 & 4.000 & 5.079 & 8.373 \\
cost & 11.99 & 48.33 & 63.39 & 68.82 & 90.68 & 150.81 \\
\end{tabular}
\end{table}
\vspace{0.5cm}

\begin{figure}
\center{\includegraphics[width=6in]{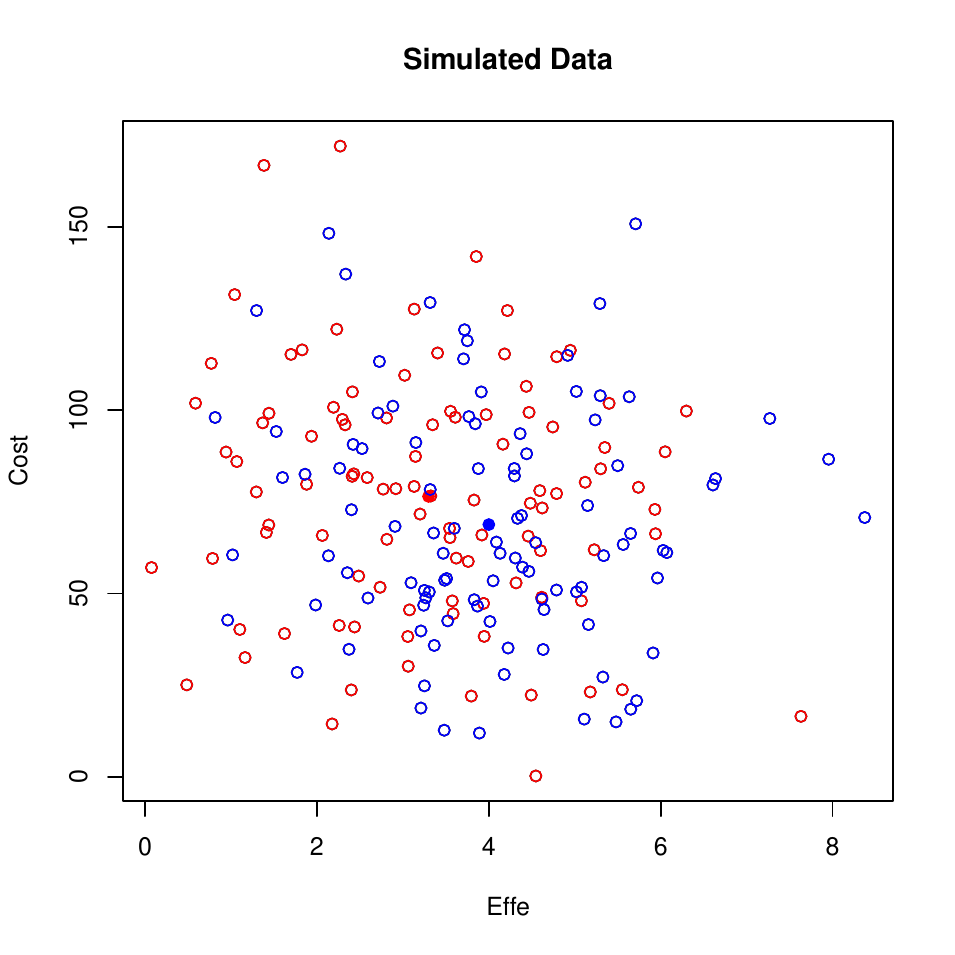}}  
\caption{\label{fig:QCdata} Simulated data on $200$ patients illustrate the case where the \textit{Shadow Price of Health}
is taken to be $\lambda = 10$. The dataset contains $99$ patients (colored Red) who received the ``Standard Treatment'' and
$101$ patients (colored Blue) who received the ``New Treatment.'' The solid Red and Blue points correspond to the two sample \textbf{Means}
of $x = Effe$ and $y = Cost$ coordinates. While these two mean-values suggest that ``New'' could be both less costly and more effective
than the ``Std'' treatment, our objective will be to estimate the statistical confidence that is associated with this possibility.}
\end{figure}

\subsection{Wedge Shaped ICE Confidence Regions}

The \textit{ICE Statistical Inferences} of interest concern \textit{Two Distinct Dimensions}: Cost and Effectiveness. Furthermore, two
(or more) treatments are to be compared within both of these dimensions; see Laupacis, Feeny, Detsky and Tugwell (1992).  Early proponents
of using polar coordinates in ICE calculations were Obenchain (1997) as well as Cook and Heyse (2000). To this day, most ``Health Economists''
talk only about \textit{Ratios} and Cartesian (rectangular) coordinates.

The first ``Wedge Shaped'' Confidence Regions were strictly \textit{parametric} regions that assumed observed data followed a
Bivariate Normal distribution and used approximations to Fieller's Theorem (1954). The accuracy of this approach was greatly improved by 
Chaudhary and Stearns (1996), and it was subsequently found that these regions are frequently \textit{highly similar} to those produced by
Bootstrapping.  

Typical analyses examine treatment differences of the form $(New - Std)$, which can be viewed as placing the $Std-$treatment at the
``ICE Origin'', $(0,0)$, shown as the Solid Red Dot in each panel of Figure $3$. The $(New - Std)$ difference is then represented by the
Solid Blue points in each panel of Figure $3$. Several properties of the ``Wedge Shaped'' (Bootstrap) ICE Confidence Regions are
pointed out in the detailed captions of Figures $3$ and $4$.
 
\begin{figure}
\center{\includegraphics[width=6in]{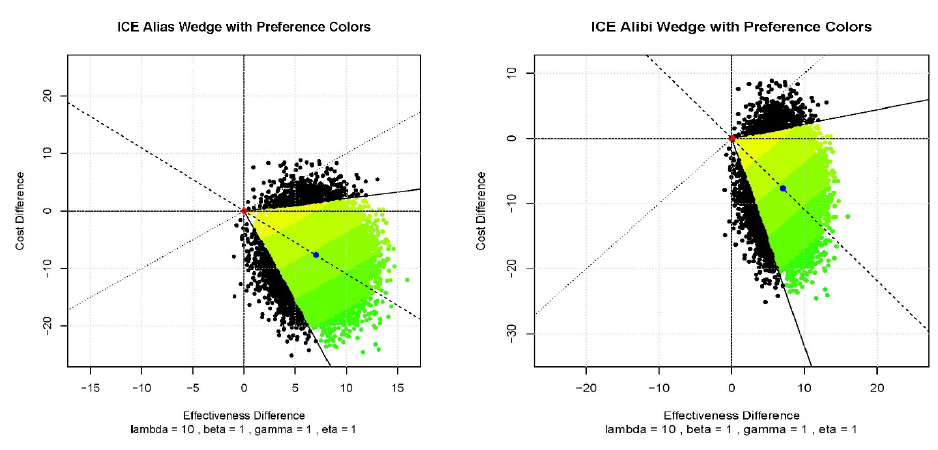}}  
\caption{\label{fig:ICEwdgNB} Although the two alternative ICE perspectives (alias and alibi) depicted here typically scale the Effe and
Cost axes differently, these perspectives still have much in common. Note that the Solid Red point in Figure $2$ appears at the ICE origin in
each of these plots, while the Solid Blue ``New'' treatment means appear near the numerical \textit{Centroid} of each Bootstrap scatter. Close examination of the points near the edges of these somewhat ``elliptical'' scatters confirms that both depict the \textit{same bootstrap scatter} containing $R = 25000$ replications. Since $688$ ICE Angle order-statistics (Black Dots) are (clockwise) below the ``lower limit'' of the \textit{Central Wedge} and $562$ are (counter-clockwise) above the Wedge ``upper limit'', each wedge contains $25000 - 688 - 562 = 23750$ replications. The \textit{Confidence Level} of both of these wedges is thus $95\%$.}
\end{figure}

\begin{figure}
\center{\includegraphics[width=6in]{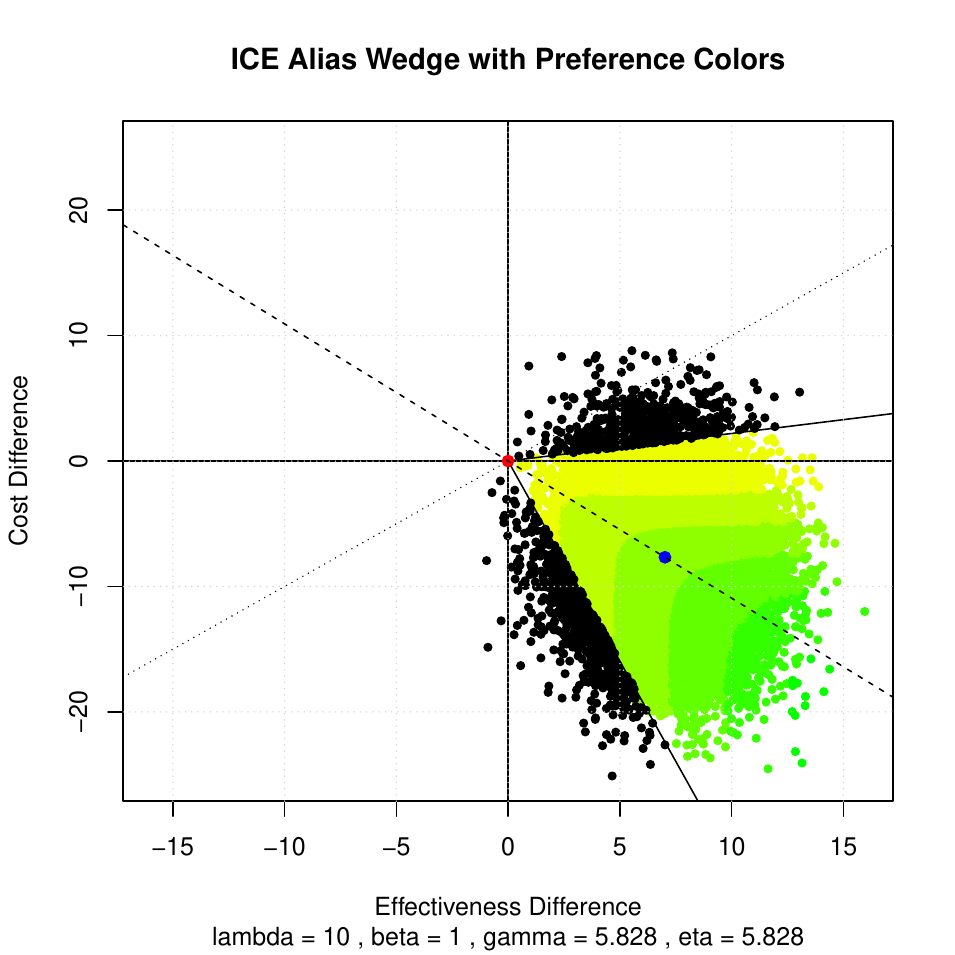}}  
\caption{\label{fig:ICEwdgOM} The $95\%$ ICE Alias \textit{Confidence Wedge} of Figure $3$ is ``Re-Colored'' here using the \textit{ICE Omega}
Preference Map that is as ``nonlinear'' as is possible and yet still satisfy \textit{Axiom Two: Monotonicity}. The ICE Alias perspective in
this (large) plot is identical to that of the left-hand sub-plot in Figure $3$. Note that there are \textit{many more numerically small} preferences (colored Yellow) within this $95\%$ Confidence Wedge than within the same wedge on Figure $3$.}
\end{figure}

\subsection{The Cost-Effectiveness Frontier}

Our example a ``Cost-Effectiveness Frontier'', depicted below in Figure $5$, is based upon one published by the
\textit{Health Economics Resource Center, U.S. Veterans Affairs.gov}. I have added a point labeled ``WW'' with
\textit{effe} $= 0$ and \textit{cost} $= 0$ that represents a hypothetical treatment strategy called \textit{Watchful Waiting}.

\begin{figure}
\center{\includegraphics[width=6in]{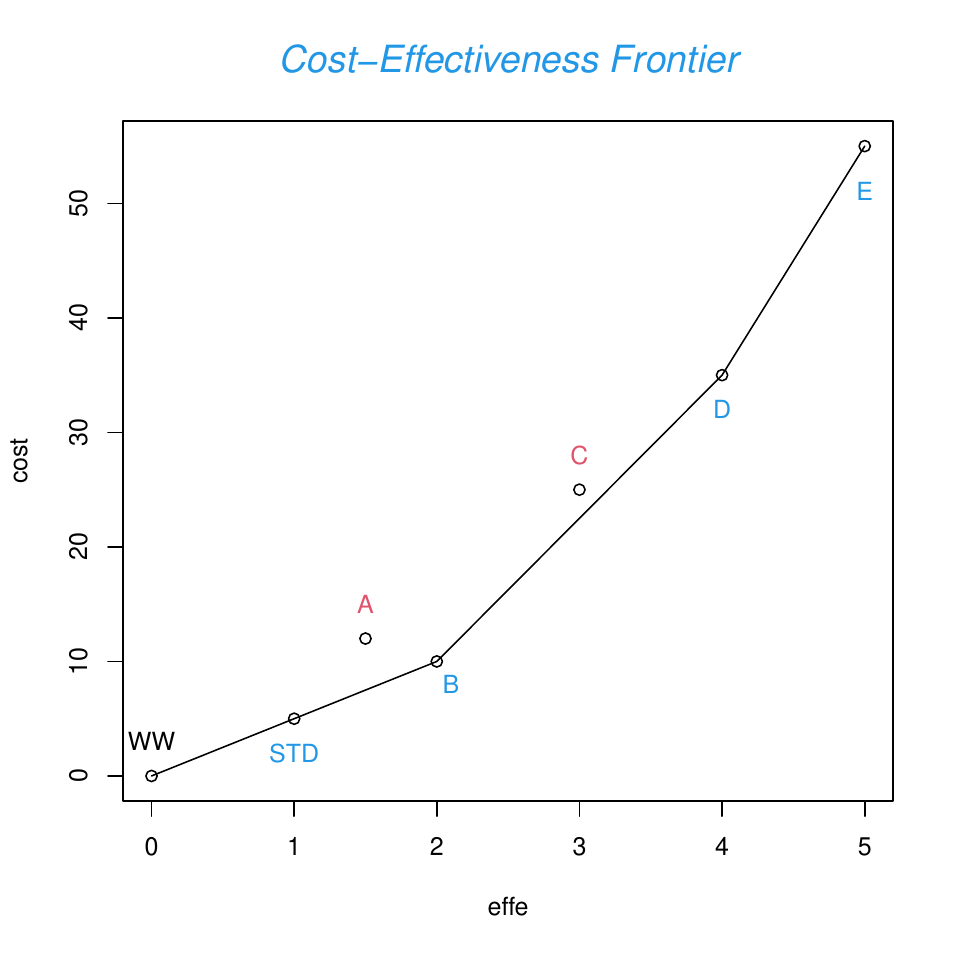}}
\caption{\label{fig:CEfront} Treatment option \textbf{A} on this graphic is (strictly) dominated by \textbf{B},
which is both \textit{more effective} and \textit{less costly} than \textbf{A}. Next, note that while treatment \textbf{C} is
more effective than \textbf{B} and less costly than \textbf{D}, treatment \textbf{C} is strictly above the straight line
segment connecting treatment options \textbf{B} and \textbf{D}. Thus, treatment \textbf{C} is said to be \textit{Extendedly Dominated}
by \textit{some mixture} of treatment options \textbf{B} and \textbf{D}. The $5$ treatment options actually on the
``Cost-Effectiveness Frontier'' are thus: \textbf{WW, STD, B, D} and \textbf{E}.}
\end{figure}

If data on measures of Effectiveness and Cost are available on patients receiving treatments \textbf{B}, \textbf{C} or \textbf{D},
functions within the \textit{ICEinfer} R-package could be used to compare outcomes on \textbf{C} with any ``Mixture'' of patients
receiving treatments \textbf{B} or \textbf{D}. Many such ``Mixtures'' may well prove to be more Cost-Effective than treatment
with option \textbf{C}.
 
\section{Recommendations}

The right-hand sides of Figures $6$ and $7$ on pages $12$ and $13$ illustrate the ICE Statistical Inference perspectives championed
here. Numerous consumer research studies have reported that preferences typically are distinctly \textit{nonlinear} ...as depicted
in the right-hand ``map'' of Figure $6$. When communicating study findings, we recommend following the ``KISS Principle'' of Zellner (1991,
2001): \textit{Keep It Sophisticatedly Simple}! 

\begin{figure}
\center{\includegraphics[width=5in]{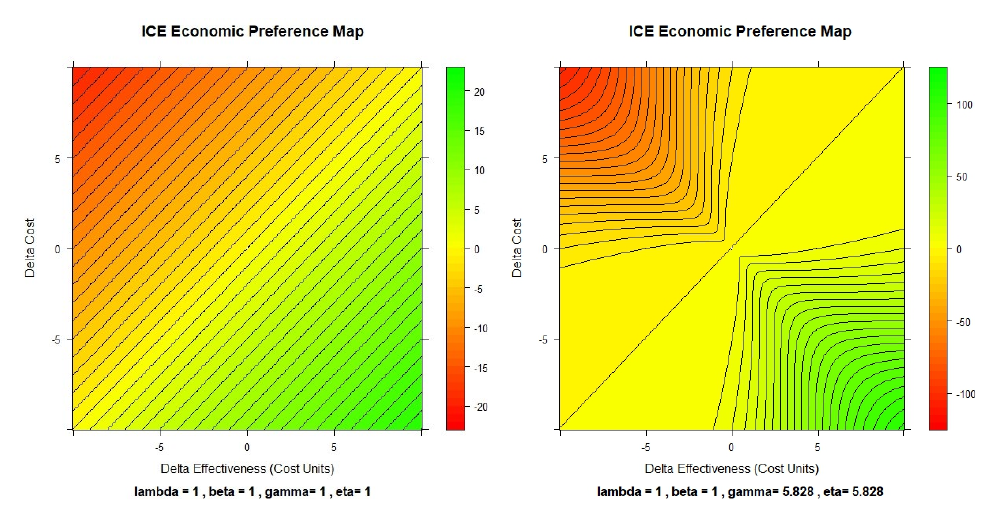}} 
\caption{\label{fig:L01maps} When $\lambda = 1$ and $\beta = 1$ (linear returns), note that there are numerous clear differences between
the \textit{ICE Preference Maps} for traditional (linear)``Net Benefit'' ($\gamma = \eta = 1$) on the Left and the (highly nonlinear)
``ICE Omega'' map on the Right ($\gamma = \eta = 5.828$). Note that both Maps are colored Yellow towards Green (numerically positive
values) below their lower-left to upper-right diagonal, and Tan towards Red (numerically negative values) above that same diagonal.   
In fact, note also that the ``Net Benefit'' map is essentially ``One - Dimensional'' (i.e. color depends only on the scalar
\textit{Effe-Cost difference}) while the nonlinear map is fully ``Two - Dimensional''.}
\end{figure}

Both histograms illustrated in Figure $7$ are highly favorable to use of the ``New'' treatment over the ``Std'' ...i.e. both
contain only \textit{strictly positive estimates} of preference for ``New''. While the left-hand histogram suggests stronger (more
positive) preferences, this less-sophisticated (linear) analysis could easily be ``badly biased''. While I recommend presenting
both histograms to all interested parties, the more-sophisticated (nonlinear) analysis is both more conservative and potentially
more realistic. In fact, this sort of non-linearity strikes me as being more likely to be confirmed by evidence from actual practice. 

\begin{figure}
\center{\includegraphics[width=6in]{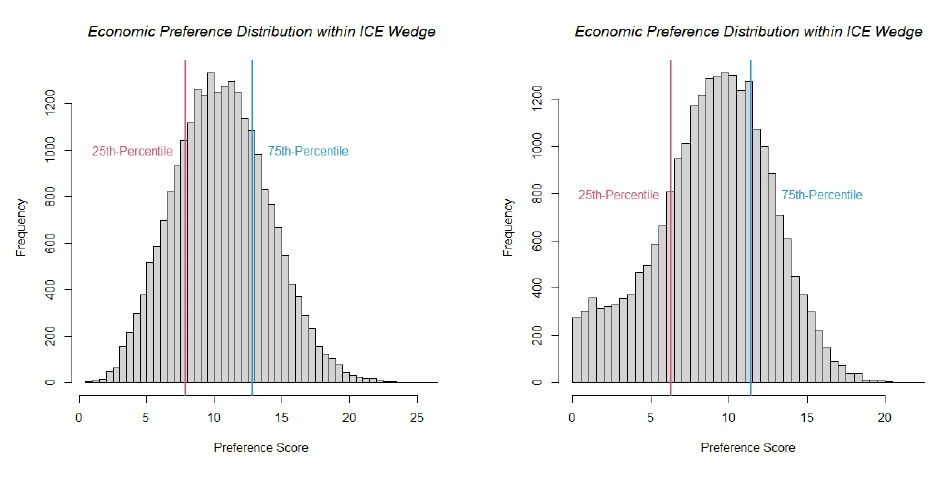}}
\caption{\label{fig:NLBhist} These two simple Histograms illustrate both similarities and differences between the potentially
over-simplistic ``Net Benefit'' evaluation on the Left and the potentially more realistic (and distinctly nonlinear) ICE Omega Preference
Map (shown in Figure $4$) on the Right. The pure simplicity of these graphics may make them ideal for convincing Health-Care administrators
and regulators that the ``New'' Treatment delivers outcomes with superior Cost-Effectiveness relative to the ``Std'' Treatment. Since
the Right-Hand Histogram quantifies Non-Linear preferences, it probably portrays a slightly more realistic (less optimistic) evaluation.}
\end{figure}

\section{Summary}

This paper has outlined and updated the statistical methodology needed by Health-Outcomes researchers to make powerful
and robust inferences using either clinical or real-world (observational) data to make realistic Head-to-Head comparisons of cost and
effectiveness differences between a pair of alternative treatments for a given disease or health-care condition.
The Cost-Effectiveness ``ICE Preference Map'' used in this approach can be distinctly non-linear to incorporate truly
realistic consumer preferences from empirical health care research. 

Over the last 50 years, computers have helped shape statistical theory as well as its practice. Freely available software can provide computational and visual fast-tracks into the strengths and weaknesses of alternative statistical methods. Why not analyze hundreds or
thousands of numbers using millions or billions of calculations, Efron (1979). In reality, \textbf{R} functions, R-Core Team (2023), are
almost indispensable tools for today's students, teachers, applied researchers and data-scientists who use and/or extend existing
computational and visualization methodologies.

\end{document}